\documentclass[twocolumn,showpacs,amsmath,amssymb,superscriptaddress,nofootinbib]{revtex4}

\newcommand{\eps}{\varepsilon}
\newcommand{\tens}[1]{{\boldsymbol{#1}}}
\newcommand{\cl}{p}
\newcommand{\cq}{c}
\newcommand{\cp}{C}
\newcommand{\cKY}{h}

\newcommand{\pder}{{\tens{\partial}}}
\newcommand{\grad}{{\tens{d}}}
\newcommand{\lied}{\pounds}
\newcommand{\covd}{{\tens{\nabla}}}
\newcommand{\coord}{{\tens{\eth}}}
\newcommand{\cv}{{\tens{\partial}}}
\newcommand{\eb}{{\tens{e}}}
\newcommand{\phsp}{{\boldsymbol{\Gamma}}}
\newcommand{\coTB}{{\mathbf{T}^{*}M}}

\DeclareMathOperator{\tr}{\mathrm{tr}}
\DeclareMathOperator{\sign}{\mathrm{sign}}

\begin{document}
\title{Constants of Geodesic Motion in Higher-Dimensional Black-Hole Spacetimes}

\author{Pavel Krtou\v{s}}

\email{Pavel.Krtous@mff.cuni.cz}

\affiliation{Institute of Theoretical Physics, Charles University, V
Hole\v{s}ovi\v{c}k\'ach 2, Prague, Czech Republic}

\author{David Kubiz\v n\'ak}

\email{kubiznak@phys.ualberta.ca}

\affiliation{Theoretical Physics Institute, University of Alberta, Edmonton,
Alberta, Canada T6G 2G7}

\affiliation{Institute of Theoretical Physics, Charles University, V
Hole\v{s}ovi\v{c}k\'ach 2, Prague, Czech Republic}

\author{Don N. Page}

\email{don@phys.ualberta.ca}

\affiliation{Theoretical Physics Institute, University of Alberta, Edmonton,
Alberta, Canada T6G 2G7}

\author{Muraari Vasudevan}

\email{mvasudev@phys.ualberta.ca}

\affiliation{Theoretical Physics Institute, University of Alberta, Edmonton,
Alberta, Canada T6G 2G7}

\affiliation{JLR Engineering, 111 SE Everett Mall Way, E-201, Everett, WA
98208-3236, USA}

\date{June 28, 2007} 
\preprint{Alberta-Thy-05-07}

\begin{abstract}
In \cite{PKVK} we announced the complete integrability of geodesic motion  in
the general higher-dimensional rotating black-hole spacetimes. In the present
paper we prove all the necessary steps leading to this conclusion. In
particular, we demonstrate the independence of the constants of motion and the
fact that they Poisson commute. The relation to a different set of constants of
motion constructed in \cite{KKPF} is also briefly discussed.
\end{abstract}

\pacs{04.70.Bw, 04.50.+h, 04.20.Jb \hfill  Alberta-Thy-05-07}    

\maketitle

\section{Introduction}
\label{sc:intro}

Spacetimes of higher dimensions (${D>4}$) have become much studied as a result of
their appearance in theories of unification, such as string/M theory.  Of such
spacetimes, one important class is a sequence of black-hole metrics of greater
and greater generality in higher dimensions that have been discovered over the
years.

The first such higher-dimensional black-hole spacetime was the metric
for a nonrotating black hole in ${D>4}$ (the generalization of the 1916
Schwarzschild metric in four dimensions \cite{Schw}), found in 1963 by
Tangherlini \cite{Tang}.  Next was the metric for a rotating black hole
in higher dimensions (the generalization of the 1963 Kerr metric in
four dimensions \cite{Kerr}), discovered in 1986 by Myers and Perry
\cite{MP} in the case with zero cosmological constant.  Then in 1998
Hawking, Hunter, and Taylor-Robinson \cite{HHT} found the general
${D\!=\!5}$ version of the ${D\!=\!4}$ rotating black hole with a
cosmological constant (often called the Kerr--\mbox{(anti-)}de~Sitter
metric) that had been found in 1968 by Carter \cite{Carter1,Carter2}. 
In 2004 Gibbons, L\"u, Page, and Pope \cite{GLPP1,GLPP2} discovered the
general Kerr--de Sitter metrics in all higher dimensions, and in 2006
Chen, L\"u, and Pope \cite{CLP} put these into a simple form similar to
Carter's and  were able to add a NUT \cite{NUT} parameter (though not
charge) to get the general  Kerr--NUT--(A)dS metrics for all $D$.

It is important to study the properties of these higher-dimensional black-hole
spacetimes, and one key property is the nature of geodesic motion in them.  In
\cite{PKVK,KKPF} we exhibited $D$ constants of geodesic motion and announced
that they are all independent (making the geodesic motion integrable) and that
the Poisson brackets of any pair of them vanish (making the integrable geodesic
motion completely integrable).  In this paper we shall prove these assertions.

After introducing the metric and its basic symmetries, we recapitulate our construction
of constants of geodesic motion and show how these constants can be generated from a generating
function. The two main proofs demonstrating the independence
and the Poisson commutativity of these constants follow. The canonical formalism
used in the text is reviewed in the Appendix.
We type tensors in boldface with
components in normal letters. 
The spacetime indices are denoted by Latin letters from the 
beginning of the alphabet, ${a,b,c=1,\dots,D}$, and
we use the Einstein summation convention for them.
For a rank-2 tensor ${\tens{B}}$ 
the symbol ${B}$ stands for the matrix 
of its components ${B^{a}_{\ b}}$.
Where it cannot lead to a confusion 
a dot indicates contraction, i.e.,
${\tens{a}\cdot\tens{b}=a^e b_e}$. 
We assume automatic lowering and raising of indices 
using the metric.
${\cv_{x^a}}$ stands for the coordinate vector 
associated with the coordinate~${x^a}$.

\section{Higher-dimensional black-hole spacetimes}
\label{sc:HDBH}
The general Kerr--NUT--(anti-)de~Sitter spacetime discovered by 
Chen, L\"u, and Pope
may, after a suitable Wick rotation of 
the radial coordinate, be written \cite{CLP} 
\begin{equation}\label{metric_coordinates}
\tens{g}\!\!=\!\!\sum_{\mu=1}^n\Bigl[\frac{\grad x_{\mu}^2}{Q_{\mu}}
  +Q_{\mu}\!\Bigl(\sum_{k=0}^{n-1} A_{\mu}^{(k)}\grad\psi_k\!\Bigr)^{\!2}\Bigr]
  \!-\!\frac{\eps c}{A^{(n)}}\Bigl(\sum_{k=0}^n A^{(k)}\grad\psi_k\!\Bigr)^{\!2} \!,
\end{equation}
with ${n=\lfloor D/2\rfloor}$ and ${\varepsilon=D-2n}$.
Here, $Q_{\mu}={X_{\mu}/}{U_{\mu}}\,,$
\begin{gather} 
U_{\mu}=\prod_{\substack{\nu=1\\\nu\ne\mu}}^{n}(x_{\nu}^2-x_{\mu}^2)\;,\;\;\;
X_{\mu}=\sum\limits_{k=\varepsilon}^{n}c_kx_{\mu}^{2k}-2b_{\mu}x_{\mu}^{1-\varepsilon}+\frac{\varepsilon c}{x_{\mu}^2}\;,
 \nonumber\\ 
A_{\mu}^{(k)}=\!\!\!\!\!\sum_{\substack{\nu_1<\dots<\nu_k\\\nu_i\ne\mu}}\!\!\!\!\!x^2_{\nu_1}\dots x^2_{\nu_k}\;,\quad 
A^{(k)}=\!\!\!\!\!\sum_{\nu_1<\dots<\nu_k}\!\!\!\!\!x^2_{\nu_1}\dots x^2_{\nu_k}\;\label{co}.
\end{gather}
The coordinates ${x_\mu}$ (${\mu=1,\dots,n}$) correspond to radial and latitude directions, 
${\psi_k}$ ( ${k=0,\dots,n-1+\eps}$) to temporal and azimuthal directions.
The parameter $c_n$ is proportional to the cosmological constant, and the
remaining constants $c_k$, $c$ and $b_{\mu}$ are 
related to the rotation parameters, the mass and the NUT parameters.
Hamamoto, Houri, Oota and Yasui 
\cite{KNAcurv} derived explicit formulas for the curvature and demonstrated that
in all dimensions this metric obeys the Einstein equations
\begin{equation}
{R_{ab}=(-1)^{n}(D-1)c_n\, g_{ab}}\;.
\end{equation}

Besides the obvious spacetime isometries generated by the $D-n$ Killing vectors
$\tens{\partial}_{\psi_k}$, the spacetime possesses
a whole set of hidden symmetries \cite{PKVK, KKPF}, which can be generated
from the principal (rank-2 closed) conformal Killing--Yano tensor discovered by Kubiz\v n\' ak and Frolov \cite{KF}.
These hidden symmetries play the crucial role for the integrability of the geodesic motion.  

The metric (\ref{metric_coordinates}) can be diagonalized. 
Let us 
introduce the orthonormal basis one-forms
\begin{eqnarray}\label{one-forms}           
\eb^\mu &=& Q_{\mu}^{-1/2} dx_{\mu}\;, \nonumber\\ 
\eb^{\hat \mu}&=&\eb^{n+\mu} = Q_{\mu}^{1/2}
 \sum_{k=0}^{n-1}A_{\mu}^{(k)}d\psi_k\;, \nonumber\\
\eb^{2n+1} &=& (-c/A^{(n)})^{1/2}
\sum_{k=0}^nA^{(k)}d\psi_k\;.
\end{eqnarray}
Then we have 
\begin{equation}\label{metric}
\tens{g}=\sum_{a=1}^D\eb^a \eb^a=\sum_{\mu=1}^n \bigl(\eb^\mu\eb^\mu 
   + \eb^{\hat\mu}\eb^{\hat\mu}\bigr)
   + \varepsilon\, \eb^{2n+1}\eb^{2n+1}\;,
\end{equation} 
and the principal conformal Killing--Yano tensor $\tens{h}$
which obeys the equations
\begin{equation}\label{cKYprop}
(D-1)\nabla_{\!a} \cKY_{bc}=g_{ab}\xi_c-g_{ac}\xi_b\;,\ \  
 \xi_a = \nabla_{\!c}\cKY^c{}_a\;,
\end{equation}
takes the extremely simple form
\begin{equation}\label{cKYdef}
\tens{\cKY}=\sum_{\mu=1}^n x_\mu \eb^\mu \wedge \eb^{\hat\mu}\;.
\end{equation}

In what follows we shall also use the conformal Killing tensor
\begin{equation}\label{Qdef}
Q=-\cKY\cKY\;,\quad\text{i.e.,}\quad Q_{ab}=\cKY_{ac}\cKY_{bd}\,g^{cd}\;,
\end{equation}
which takes the explicit form
\begin{equation}\label{Qexpl}
\tens{Q}=\sum_{\mu=1}^nx_{\mu}^2(\eb^\mu\eb^\mu + \eb^{\hat\mu}\eb^{\hat\mu})\;,
\end{equation} 
and satisfies $\nabla_{\!(a} Q_{bc)}=g_{(ab}Q_{c)}\,,$ where
\begin{equation}
Q_a =\frac{1}{D+2}(2\nabla_{\!c}\,Q^c_{\ a}+\nabla_{\!a}Q^c_{\ c}).
\end{equation}

\section{Constants of motion}
\label{sc:cm}

In \cite{PKVK} we have claimed that in the spacetime \eqref{metric}
there are ${D}$ independent constants of geodesic motion, given by the 
following quantities: 
(a) $n-1$ observables $\cp_j$, ${j=1,\dots,n-1}$, given by
traces of powers of the projection $\tens{F}$ of the 
principal conformal Killing-Yano tensors~$\tens{\cKY}$ (cf.\ Eqs.~\eqref{Fdef} and \eqref{F=PKP} below)
\begin{equation}\label{cpdef}
  \cp_j = \tr\bigl[(-w^{-1}F^2)^j\bigr]\;,
\end{equation}
(b) ${D-n}$ observables ${\cl_j}$, ${j=0,\dots,D-n-1}$, 
given by symmetries of the spacetime 
\begin{equation}\label{cldef}
\cl_j=\tens{u}\cdot\cv_{\psi_j}\;,
\end{equation}
and (c) the square ${w}$ of the (unnormalized) velocity ${\tens{u}}$
\begin{equation}\label{wdef}
w=\tens{u}\cdot\tens{u}=u^a u_a\;.
\end{equation}
Moreover, these quantities commute in the sense of Poisson brackets on the phase space.
Here we want to elucidate and prove these properties in more detail.

We understand all mentioned quantities as observables
(i.e., functions) on the phase space ${\phsp=\coTB}$.
It is well known that the cotangent space ${\coTB}$ has a natural phase space
structure (cf. the Appendix or \cite{Arnold:book}).
Since we investigate the relativistic theory 
and ${M}$ is a spacetime manifold describing also the physical
temporal direction, the phase space ${\phsp=\coTB}$
is an unphysical phase space which is, however, well suited for an investigation of the 
geodesic motion. Doing canonical mechanics on it allows us to solve
the geodesic motion in an external time which can be identified
at the end with the affine parameter of the studied geodesic.

We denote the momentum variable on the cotangent space 
as ${\tens{u}}$. Indeed, since the geodesic motion is governed by
the Lagrange function ${L=\frac12\,\tens{u}\cdot\tens{u}=\frac12\, u^a u^b\, g_{ab}}$,
the canonical momentum can be (up to a position of the tensor index)
identified with the (unnormalized) velocity ${\tens{u}}$.
The Hamiltonian then is
\begin{equation}\label{Ham}
  H=\frac12\,w=\frac12\,\tens{u}\cdot\tens{u}=\frac12\,u_a u_b\, g^{ab}\;.
\end{equation}

We easily realize that ${\cl_j}$ defined in \eqref{cldef} are
the special components of momentum and that they
are constants of motion since $\cv_{\psi_j}$ are Killing vectors. 
The quantities $\cp_j$, eq.~\eqref{cpdef}, are constants of motion because 
the tensor $\tens{F}$, defined as 
\begin{equation}\label{Fdef}
F_{ab}=(\cKY_{ab}u_c+\cKY_{bc}u_a+\cKY_{ca}u_b)\;u^c ,
\end{equation}
and the square ${w}$ of the velocity ${\tens{u}}$,
are covariantly conserved along the geodesic. Indeed, thanks to \eqref{cKYprop},
for ${u^c\nabla_{\!c} u^a=0}$ we have $u^c\nabla_{\!c} F_{ab}=0$.

Next we express the constants of motion ${\cp_j}$ in terms of the
quantities related to the principal conformal Killing-Yano tensor ${\tens{\cKY}}$.
The components of the tensor \eqref{Fdef} can be rewritten as
\begin{equation}\label{F=PKP}
w^{-1}F=P\cKY P\;,
\end{equation}
where ${\tens{P}}$ is the projector orthogonal to the velocity ${\tens{u}}$,
${P=I-p}$, i.e., ${P^a_{b}=\delta^a_{b}-p^a_{b}}$.
Here we also introduced the projector ${\tens{p}}$,  
\begin{equation}\label{Pdef}
p^a_{b}=w^{-1} u^a u_b\;,
\end{equation}
onto the direction ${\tens{u}}$. Using the cyclic property of the trace we thus have
\begin{equation}\label{Cexpr}
\cp_j=(-1)^j w^j \tr\bigl[(\cKY P)^{2j}\bigr]\;.
\end{equation}
The trace of the matrix product could be viewed diagrammatically as a loop
formed by joined vertices (each with two `legs') corresponding to matrices 
in the product. In our case the loop is formed by alternating ${\cKY}$ and ${P}$
vertices. Substituting ${P=I-p}$ we get a sum over all possible loops
in which ${P}$ is replaced either by ${I}$ or by ${-p}$. In the 
case of the identity ${I}$ the corresponding vertex is effectively
eliminated, and in the case of the one-dimensional projector 
${\tens{p}=w^{-1}\tens{u}\tens{u}}$ the loop splits into disconnected pieces.
Namely, we can use the identity
\begin{equation}
\tr\bigl(\cKY^{k_1}\!p\,\cKY^{k_2}\!p\,\cdots\cKY^{k_c}\!p\bigr)=
\tr\bigl(\cKY^{k_1}\! p\bigr)\tr\bigl(\cKY^{k_2}\! p\bigr)\cdots\tr\bigl(\cKY^{k_c}\! p\bigr)\;.
\end{equation}

The trace in \eqref{Cexpr} thus leads to
\begin{equation}\label{trhPj}
\begin{split}
&\tr\bigl[(\cKY P)^{2j}\bigr]=\tr\bigl(\cKY^{2j}\bigr)\\
  &\qquad+\sum_{c=1}^{2j}\sum_{\substack{k_1\le \dots\le k_c\\k_1+\dots+k_c=2j}}
  \!\!\!(-1)^c N^{2j}_{k_1\dots k_j} \prod_{i=1}^c\tr\bigl(\cKY^{k_i}\!p\bigr)\;.
\end{split}
\end{equation}
The sum over ${c}$ is the sum over the number of `splits' of the loop,
the indices ${k_i}$ are the `lengths' of the split pieces,
and the combinatorial factor ${N^{2j}_{k_1\dots k_c}}$ 
gives the number of ways in which the loop of the length ${2j}$
can be split to ${c}$ pieces of lengths ${k_1,\dots,k_c}$.
 From the fact that the tensor ${\tens{\cKY}}$ is antisymmetric, 
it follows that traces of odd powers of ${\cKY}$ (optionally multiplied by 
a projector) are zero. Setting ${k_i=2l_i}$ 
and introducing the rank-2 conformal Killing tensor ${\tens{Q}}$ from \eqref{Qdef},
eq.~\eqref{trhPj} thus reduces to
\begin{equation}\label{trhPjinQ}
\begin{split}
&\tr\bigl[(-\cKY P\cKY P)^{j}\bigr]=\tr\bigl(Q^j\bigr)\\
  &\qquad+\sum_{c=1}^j\sum_{\substack{l_1\le \dots\le l_c\\l_1+\dots+l_c=j}}
  \!\!\!(-1)^c\, 2\,N^j_{l_1\dots l_j} \prod_{i=1}^c\tr\bigl(Q^{l_i}\!p\bigr)\;,
\end{split}
\end{equation}
where we used ${N^{2j}_{2l_1\dots2l_c}=2N^j_{l_1\dots l_c}}$ 
which follows from the definition of the ${N}$'s.
If we define the quantities
\begin{equation}\label{wjdef}
  w_j = w \tr(Q^j p) = u_{a_0} Q^{a_0}_{a_1}Q^{a_1}_{a_2}\cdots
  Q^{a_{j-1}}_{a_j}u^{a_j},
\end{equation}
we finally obtain
\begin{equation}\label{trhPjinw}
\begin{split}
\cp_j&=w^j\tr\bigl[(-\cKY P\cKY P)^{j}\bigr]\\
  &=w^j\tr\bigl(Q^j\bigr)
  +2\sum_{c=1}^j\sum_{\substack{l_1\le \dots\le l_c\\l_1+\dots+l_c=j}}
  \!\!\!(-1)^c \,N^j_{l_1\dots l_j} w^{j-c}\prod_{i=1}^c w_{l_i}\;,
\end{split}
\end{equation}
which is eq. (17) of \cite{PKVK}.

Let us note that by the same argument as that leading to eq.~\eqref{trhPj}, 
we can derive the relation for the trace of a power of ${QP}$,
\begin{equation}\label{trQPj}
\begin{split}
&\tr\bigl[(Q P)^j\bigr]=\tr\bigl(Q^j\bigr)\\
  &\qquad+\sum_{c=1}^j\sum_{\substack{l_1\le \dots\le l_c\\l_1+\dots+l_c=j}}
  \!\!\!(-1)^c N^j_{l_1\dots l_j} \prod_{i=1}^c\tr\bigl(Q^{l_i}\!p\bigr)\;,
\end{split}
\end{equation}
Comparing with eq.~\eqref{trhPjinQ}, we see that we have proved
the relation (16) of \cite{PKVK},
\begin{equation}\label{trtrtr}
\tr\bigl[(-\cKY P\cKY P)^{j}\bigr]+\tr\bigl[Q^j\bigr]=2\tr\bigl[(Q P)^j\bigr]\;.
\end{equation}

The relation \eqref{trhPjinw} and an algorithm for computing 
the coefficients ${N^j_{l_1\dots l_c}}$ can be derived also in a different way.
It was mentioned in \cite{KKPF} that the constants ${\cp_j}$ can be generated
from the generating function ${Z(\beta)=\log W(\beta)}$:
\begin{equation}
\begin{split}
Z(\beta)&=\sum_{j=1}^\infty\frac{(-1)^{j+1}}{2j}\frac{\beta^j}{w^j}\cp_j
        =-\sum_{j=1}^\infty\frac{1}{2j}\beta^j\tr\bigl[(\cKY P)^{2j}\bigr]\\
        &=\tr\log\bigl(I\!-\!\sqrt\beta\,\cKY P\bigr)
        =\log\det{}\bigl(I\!-\!\sqrt\beta\,\cKY P\bigr)\;.
\end{split}
\end{equation}
The third equality follows from the antisymmetry of ${\cKY}$.
Using properties of the determinant, the antisymmetry of ${\cKY}$, 
${I=P+p}$, and the fact that the projector ${p}$
is one dimensional, we can split ${Z(\beta)}$ into two pieces 
(cf.\ eq.~(2.7) and (2.8) of \cite{KKPF}):
\begin{equation}
\begin{aligned}
Z(\beta) &= \log W_0(\beta) + \log\Sigma(\beta)\;,\\
W_0(\beta)&=\det\bigl(I\!-\!\sqrt\beta\,\cKY\bigr) 
  =\det{}^{\!\frac12\!}\bigl(I\!+\!\beta\,Q\bigr)\;,\\
\Sigma(\beta)&=\det\bigl(P+(I-\sqrt\beta\,\cKY)^{\!-\!1}p\bigr)\\
&=\tr\bigl((I-\sqrt\beta\,\cKY)^{\!-\!1}p\bigr)
=\tr\bigl((I+\beta\,Q)^{\!-\!1}p\bigr)\;.
\end{aligned}
\end{equation}

Equation \eqref{trhPjinw} then corresponds to the term
proportional to ${\beta^j}$ in the power expansion of ${Z(\beta)}$.
The first term of \eqref{trhPjinw} is obtained from ${\log W_0(\beta)}$,
and the sum over all possible splittings of the loop corresponds 
to the ${\beta^j}$ term of ${\log\Sigma(\beta)}$. Clearly, the ${j}$-th derivative 
of ${\log\Sigma(\beta)}$ (evaluated at ${\beta=0}$) contains 
the sum over all possible products of ${l}$-th derivatives ${\Sigma^{(l)}(0)}$
which are proportional to ${w_l}$ defined in \eqref{wjdef}. 
The factors ${N^j_{l_1\dots l_2}}$ can thus be obtained by the explicit
computation of the derivatives of the generating function ${\log\Sigma(\beta)}$:
\begin{equation}\label{Ck}
\begin{split}
\cp_j&=w^j\tr\bigl(Q^j\bigr)\\
  &-\frac{2(-w)^{j}}{(j-1)!}\frac{d^j}{d\beta^j}\log
\Bigr(1+\sum_{k=1}^j(-1)^k\frac{w_k}{w}\beta^k\Bigl)\Big|_{\beta=0}\;.
\end{split}
\end{equation}

Using software for algebraic manipulation we easily get 
the first five constants (sufficient for the integrability of
geodesic motion up through $D=13$):
\begin{align}
  \cp_1 &= w \tr Q - 2 w_1\;,\notag\\
  \cp_2 &= w^2 \tr Q^2 - 4 w\, w_2 + 2 w_1^2\;,\notag\\ 
  \cp_3 &= w^3 \tr Q^3 - 6 w^2 w_3 + 6 w\, w_1 w_2 - 2 w_1^3\;,\notag\\
  \cp_4 &= w^4 \tr Q^4 - 8 w^3 w_4 + w^2 (4w_2^2+8w_1w_3)\label{c1-5}\\
        &\qquad-8 w\, w_1^2w_2+2w_1^4\;,\notag\\ 
  \cp_5 &= w^5 \tr Q^5 - 10 w^4 w_5 + w^3 (10w_2w_3+10w_1w_4)\notag\\
        &\qquad-w^2(10w_1w_2^2+10w_1^2w_3)+10w\,w_1^3w_2-2w_1^5\;.\notag
\end{align}

Taking into account the facts that the eigenvalues of the principal conformal 
Killing-Yano tensor ${\tens{\cKY}}$ are given by the coordinates ${x_\mu}$, 
cf.\ eq.~\eqref{cKYdef}, respectively, that the eigenvalues of ${\tens{Q}}$
are ${x_\mu^2}$, see eq.~\eqref{Qexpl}, we can write down an explicit form for ${\tr Q^j}$ and ${w_j}$:
\begin{equation}\label{trQjinx}
  \tr Q^j = 2 \sum_{\mu=1}^n x_\mu^{2j}\;,
\end{equation}
\begin{equation}\label{wjinx}
  w_j = \sum_{\mu=1}^n x_\mu^{2j} (u_\mu^2+u_{\hat\mu}^2)\;.
\end{equation}

Let us also point out that on the level of the generating 
functions the relation \eqref{trtrtr} corresponds to
\begin{equation}\label{detdetdet}
  \det\bigl(I-\beta\,\cKY P\cKY P\bigr)\,\det\bigl(I+\beta\,Q\bigr)
  =\det{}^{\!2}\bigl(I+\beta\, QP\bigr)\;.
\end{equation}

It was realized in \cite{KKPF} that the generating function ${W(\beta)=\exp Z(\beta)}=
W_0(\beta)\Sigma(\beta)$
actually generates another set of conserved quantities ${\cq_j}$ by
\begin{equation}\label{quadrc}
  W(\beta) = \frac1w\sum_{j=0}^\infty \cq_j\beta^j\;,
\end{equation}
which are quadratic in the velocity~${\tens{u}}$. 
(That they are quadratic can be seen from the fact that ${W_0}$
does not depend on the velocity, from eq.~\eqref{wjinx}, 
and from ${w \Sigma(\beta) = \sum_{j=0}^\infty (-1)^j w_j \beta^j}$.)
The relation between ${W(\beta)}$ and ${Z(\beta)}$ implies that
\begin{equation}\label{Ckck}
\cp_j=-\frac{2(-w)^{j}}{(j-1)!}\frac{d^j}{d\beta^j}
\log\Bigl(w+\sum_{k=1}^jc_k\beta^k\Bigr)\Big|_{\beta=0}\;,
\end{equation}
and in particular:
\begin{align}
\cp_1&=\;\;\,2\cq_1\;,\notag\\
\cp_2&=-4w\cq_2+2\cq_1^2\;,\notag\\
\cp_3&=\;\;\,6w^2\cq_3-6w\cq_1\cq_2+2\cq_1^3\;,\label{cCrel}\\
\cp_4&=-8w^3\cq_4+8w^2\cq_1\cq_3+4w^2\cq_2^2-8w\cq_1^2\cq_2+2\cq_1^4\;,\notag\\
\cp_5&=\;10w^4\cq_5-10w^3\cq_1\cq_4-10w^3\cq_2\cq_3\notag\\
     &\qquad+10w^2\cq_1^2\cq_3+10w^2\cq_1\cq_2^2-10w\cq_1^3\cq_2+2\cq_1^5\;,\notag
\end{align}
which are the inverse of the relations (3.19) of \cite{KKPF}.

\section{Independence of constants of motion}

Now we can demonstrate that the quantities ${w}$, ${\cl_j}$  and ${\cp_j}$ are independent 
at a generic point of the phase space ${\phsp=\coTB}$. This means that 
their gradients on the phase space are linearly independent. 
To prove this, it is sufficient to show that these gradients
are independent in the vertical direction of the cotangent bundle ${\coTB}$,
i.e., that the derivatives of these quantities with respect to the momentum ${\tens{u}}$
are linearly independent. To achieve this we will study the wedge product of
the `vertical' derivatives. 

Let us, instead of ${w}$ and ${\cp_j}$, 
consider the equivalent set of observables ($j=1,\dots,n-1$)
\begin{equation}\label{cscal1}
\begin{aligned}
  2\tilde{\cp}_j&=-\frac1{2j}\,w^{1-j}\cp_j=-\frac{1}{2j}\,w\tr{{Q}^j}+w_j+\dots\;,\\
  2\tilde{\cp}_0&=w\;,
\end{aligned}
\end{equation}
where dots in the first expression denote terms which contain $w_k$ 
with $k<j$, cf.\ eqs.~\eqref{trhPjinw}, \eqref{Ck}. 


We are interested in the quantity\footnote
{
The derivative ${\pder f}$ is the \emph{vector} field on 
spacetime ${M}$ with components ${\partial{f}/\partial{u_a}}$, cf.\ the Appendix \ref{sc:apx}
(it could be written more explicitly as ${\pder f/\pder \tens{u}}$).
The wedge product is, strictly speaking, defined for (antisymmetric) \emph{forms}. 
However, we can easily define the wedge product also for the vectors or
lower the vector indices with the help of the metric to get 1-forms.
} 
\begin{equation}\label{jac}
  \tens{J} = \pder\tilde{\cp}_0\wedge\dots\wedge\pder\tilde{\cp}_{n\!-\!1}\wedge
      \pder\cl_0\wedge\dots\wedge\pder\cl_{D\!-\!n\!-\!1}\;.
\end{equation}
Due to \eqref{cldef} and \eqref{cscal1}, we have $ \pder\cl_j=\cv_{\psi_j}\,,$
and
\begin{equation}\label{pderofcj}
  \pder\tilde\cp_j=-\frac{1}{2j}\,\bigl(\tr{Q^j}\bigr)\;\tens{u}+\tens{Q}^j\cdot\tens{u}+
\dots\;,
\end{equation}
where dots denote linear combinations of ${\tens{Q}^k\cdot\tens{u}}$ with ${k<j}$; 
${\tens{Q}^l\cdot\tens{u}}$ represents the vector with components
${Q^a_{a_1}Q^{a_1}_{a_2}\cdots Q^{a_{l\!-\!1}}_{a_j}u^{a_l}}$.
 From the antisymmetry of the wedge product it follows that
\begin{equation}\label{jacinu}
  \tens{J}=\tens{u}\wedge(\tens{Q}\cdot\tens{u})\wedge\dots\wedge(\tens{Q}^{n\!-\!1}\cdot\tens{u})\wedge
  \cv_{\psi_0}\wedge\dots\wedge\cv_{\psi_{D\!-\!n\!-\!1}}\;.
\end{equation}
(Matrix) powers ${\tens{Q}^j}$ of the conformal Killing tensor can be written as
\begin{equation}\label{powQ}
  \tens{Q}^j = \sum_{\mu=1}^n x_\mu^{2j}\eb_\mu\eb^\mu+\sum_{\mu=1}^n x_\mu^{2j}\eb_{\hat\mu}\eb^{\hat\mu}\;.
\end{equation}
The second term acts on the subspace of the vectors spanned on ${\cv_{\psi_j}}$.
Thus, thanks to the ${\cv_{\psi_0}\wedge\dots\wedge\cv_{\psi_{D\!-\!n\!-\!1}}}$ term in the wedge product,
this part can be ignored in \eqref{jacinu}. 
Taking into account that ${\eb_\mu\eb^\mu = \cv_{x_\mu}\grad x_\mu}$ and ${u^\mu=\grad x_\mu\cdot\tens{u}}$, 
the substitution of \eqref{powQ} into \eqref{jacinu} leads to
\begin{equation}\label{jacfin}
  \tens{J} = u^1 \dots u^n\, U\, \cv_{x_1}\wedge\dots\wedge\cv_{x_n}\wedge\cv_{\psi_0}\wedge\dots\wedge\cv_{\psi_{D\!-\!n\!-\!1}}\;,
\end{equation}
where 
\begin{equation}\label{Udef}
  U = \mspace{-20mu}\sum_{\substack{\text{permutations ${\sigma}$}\\\text{of ${[0,\dots, n\!-\!1]}$}}}\mspace{-20mu}
  \sign\sigma\; x_1^{2\sigma_1}\dots x_n^{2\sigma_n}
    = \mspace{-10mu}\prod_{\substack{\mu,\nu=1\dots n\\\nu<\mu}}\mspace{-8mu} (x_\mu^2-x_\nu^2)\;.
\end{equation}
In a generic point of the phase space we have ${u^j\ne0}$ and ${x_\mu^2\ne x_\nu^2}$ (for ${\mu\ne\nu}$) and
therefore ${\tens{J}\ne0}$ there, thus showing that the constants of motion  are independent.

\section{Poisson brackets}
\label{sc:pb}

Finally we show that the observables ${w}$, ${\cp_j}$, and ${\cl_j}$
Poisson commute on the phase space. 

The Poisson bracket of two functions on the phase space ${\phsp=\coTB}$ can be written as
\begin{equation}\label{PBdef}
  \{A,B\}=\covd A \cdot \pder B - \pder A \cdot \covd B\;,
\end{equation}
where ${\covd F}$ represents an arbitrary (torsion-free) covariant derivative 
which ignores the dependence of ${F}$ on the momentum ${\tens{u}}$,
and ${\pder B}$ is the derivative of ${B}$ with respect to the momentum ${\tens{u}}$,
cf.\ the Appendix. ${\covd F}$ is a 1-form and ${\pder F}$ a vector field on 
the spacetime ${M}$, and the dot indicates the contraction in spacetime tensor indices. 
We use naturally the covariant derivative ${\covd}$ generated by the metric on ${M}$.

Clearly, the commutation of any observable with
the Hamiltonian ${\frac12w}$ of the geodesic motion is equivalent to 
the conservation of the observable, cf.\ eq.~\eqref{PBHam}, so we have
\begin{equation}\label{PBw}
  \{w,\cl_j\}=0\;,\quad\{w,\cp_j\}=0\;.
\end{equation}

The Poisson bracket between observables 
${\cl_j=\tens{u}\cdot\cv_{\psi_j}}$ reduces to Lie brackets of the 
Killing vector fields ${\cv_{\psi_j}}$, which vanish because 
${\cv_{\psi_j}}$ are coordinate vector fields:
\begin{equation}\label{PBclcl}
\begin{split}
  \{\cl_i,\cl_j\} 
    &= \cv_{\psi_j}\cdot(\covd\cv_{\psi_i})\cdot\tens{u}
     -\cv_{\psi_i}\cdot(\covd\cv_{\psi_j})\cdot\tens{u}\\
    &= [\cv_{\psi_j},\cv_{\psi_i}]\cdot\tens{u}=0\;
\end{split}
\end{equation}

The Poisson bracket of any observable with the observable ${p=\tens{l}\cdot\tens{u}}$ linear in momentum
leads to the Lie derivative along the vector field ${\tens{l}}$, see\ eq.~\eqref{PBlinobs}:
\begin{equation}\label{PBclcp}
\{\cp_i,\cl_j\} 
  =\lied_{\cv_{\psi_j}}\cp_i=0\;.
\end{equation}
Here, the Lie derivative ${\lied_{\cv_{\psi_j}}\cp_i}$ ignores 
the dependence of ${\cp_i}$ on the momentum ${\tens{u}}$, cf.\ the Appendix.
It vanishes because ${\cv_{\psi_j}}$ is a Killing vector 
and the definition of ${\cp_i}$ respects the symmetry of the
spacetime (it does not depend explicitly on ${\psi_j}$).

Finally, it remains to evaluate the brackets ${\{\cp_i,\cp_j\}}$.
To simplify the following computation, we will study rescaled observables\footnote{%
The scaling \eqref{cscal2} differs from \eqref{cscal1} used in the previous section.}
\begin{equation}\label{cscal2}
  \tilde\cp_j=(-1)^jw^{j}\cp_j=\tr\bigl[(\cKY\, \tilde{P})^{2j}\bigr]\;,
\end{equation}
cf.\ eq.~\eqref{Cexpr}, and we denote
\begin{equation}\label{tPdef}
  \tilde{\tens{P}}=w\tens{P}=w\tens{I}-\tens{u}\tens{u}\;.
\end{equation}
Using the cyclic property of the trace, 
the derivative of ${\tilde\cp_j}$ in the spacetime direction is
\begin{equation}
  \nabla_a\tilde\cp_j=2j\tr\bigl[(\nabla_a\cKY) \tilde{P} (\cKY \tilde{P})^{2j-1}\bigr]\;.
\end{equation}
Here ${\nabla_a\cKY}$ is the matrix of components ${\nabla_a\cKY^b{}_c}$
of the covariant derivative ${\covd\tens{\cKY}}$. Substituting for ${\nabla_a\cKY^b{}_c}$ 
from eq.~\eqref{cKYprop} and using the antisymmetry of ${\cKY}$, we obtain
\begin{equation}\label{covdcp}
\begin{split}
  &{\textstyle\frac{D-1}{2j}}\nabla_e\tilde\cp_j\\
  &\quad=\xi_{a_0}\tilde{P}^{a_0}_{\,b_1}\cKY^{b_1}{}_{\!a_1}\tilde{P}^{a_1}_{\,b_2}
         \dots\cKY^{b_{2j-2}}{}_{\!\!\!a_{2j-1}}\tilde{P}^{a_{2j-1}}_{\,e}\\
  &\qquad-g_{ea_{2j}}\tilde{P}^{a_{2j}}_{\,b_{2j-1}}\cKY^{b_{2j-1}}{}_{\!\!\!a_{2j-1}}
         \dots\cKY^{b_1}{}_{\!a_1}\tilde{P}^{a_1}_{b_0}\xi^{b_0}\\
  &\quad=2\,\xi_{a_0}\tilde{P}^{a_0}_{\,b_1}\cKY^{b_1}{}_{\!a_1}\tilde{P}^{a_1}_{\,b_2}
         \dots\cKY^{b_{2j-2}}{}_{\!\!\!a_{2j-1}}\tilde{P}^{a_{2j-1}}_{\,e}\;.
\end{split}
\end{equation}
For the derivative with respect of the momentum ${\tens{u}}$ we get
\begin{equation}\label{pdercp}
\begin{split}
  {\textstyle\frac{1}{4j}}\partial^e \tilde\cp_j
  &=u^e\; \bigl(\cKY^{d_1}{}_{\!c_1}\tilde{P}^{c_1}_{\,d_2}\cKY^{d_2}{}_{\!c_2}\tilde{P}^{c_2}_{\,d_3}\dots
     \tilde{P}^{c_{2j\!-\!1}}_{\,d_{2j}}\cKY^{d_{2j}}{}_{\!d_1}\bigr)\\
  &\quad+ \cKY^{e}{}_{\!c_1}\tilde{P}^{c_1}_{\,d_2}\cKY^{d_2}{}_{\!c_2}\tilde{P}^{c_2}_{\,d_3}\dots
     \tilde{P}^{c_{2j\!-\!1}}_{\,d_{2j}}\cKY^{d_{2j}}{}_{\!\!c_{2j}}u^{c_{2j}}\;.
\end{split}  
\end{equation}
Substituting \eqref{covdcp} and \eqref{pdercp} into expression \eqref{PBdef} for ${\{\tilde\cp_i,\tilde\cp_j\}}$
and using ${\tilde{P}^a_{\,b}u^b=0}$, we find
\begin{equation}\label{PBcpcp}
\begin{split}
  &{\textstyle\frac{D-1}{16ij}}\{\tilde\cp_i,\tilde\cp_j\}=\\
  &\quad=\xi_{a_0}\tilde{P}^{a_0}_{\,b_1}\cKY^{b_1}{}_{\!a_1}
     \dots\tilde{P}^{a_{2i-1}}_{\,b_{2i-1}} 
  \cKY^{b_{2i-1}}{}_{\!c_1}\dots
     \tilde{P}^{c_{2j-1}}_{\,d_{2j}}\cKY^{d_{2j}}{}_{\!\!c_{2j}}u^{c_{2j}}\\     
  &\quad\,-\xi_{a_0}\tilde{P}^{a_0}_{\,b_1}\cKY^{b_1}{}_{\!a_1}
     \dots\tilde{P}^{a_{2j-1}}_{\,b_{2j-1}} 
  \cKY^{b_{2j-1}}{}_{\!c_1}\dots
     \tilde{P}^{c_{2i-1}}_{\,d_{2i}}\cKY^{d_{2i}}{}_{\!\!c_{2i}}u^{c_{2i}}\\
  &\quad =0\;.
\end{split}\raisetag{3ex}          
\end{equation}

We thus proved that the conserved quantities ${w}$, ${\cl_j}$, and ${\cp_j}$
Poisson commute with each other. Since the generating function ${Z(\beta)}$
is given by power series in ${\beta}$ with coefficients
given (up to constant factors) by the constants ${\cp_j}$, 
then also this function (and similarly ${W(\beta)=\exp Z(\beta)}$)
Poisson commute with ${w}$ and ${\cl_j}$, as well as with itself
for different choices of ${\beta}$:
\begin{equation}\label{PBZZ}
  \{Z(\beta_1),Z(\beta_2)\}=0\;,\quad\{W(\beta_1),W(\beta_2)\}=0\;.
\end{equation}
The same is true also for the quantities ${\cq_j}$ generated from ${W(\beta)}$
introduced in \cite{KKPF}. Therefore the constants of motion all
Poisson commute (are in involution), so the geodesic motion is
completely integrable \cite{Arnold:book, Kozlov}.

\section{Summary}
We have explicitly proved the complete integrability of geodesic motion 
in the general higher-dimensional rotating black-hole spacetimes \cite{CLP}.
The `nontrivial' constants of motion are associated with the Killing tensors 
which we generated from the principal conformal Killing--Yano tensor.
Observables $c_j$ are quadratic in momenta and correspond to rank-2 Killing tensors, whereas 
constants $C_j$ are of higher order in momenta and correspond to Killing tensors of increasing
rank.

The complete integrability of the geodesic motion is related to the
issue of separability of the Hamilton-Jacobi equation 
recently accomplished by Frolov, Krtou\v s, and Kubiz\v n\' ak \cite{FKK}.
The relation between integrability and separability on a general level has been
studied in the series of papers by Benenti and Francaviglia (see, e.g., \cite{Benenti})
where it was demonstrated that the separability 
is possible only if all the constants of motion, corresponding to Killing vectors
and rank-2 Killing tensors, Poisson commute.

\section*{Acknowledgments}
P.K.\ is supported by  the grant GA\v{C}R 202/06/0041 and 
appreciates the hospitality of the University of Alberta. 
D.K.\ is grateful to the Golden Bell Jar Graduate
Scholarship in Physics at the University of Alberta. 
D.P.\ thanks the Natural Sciences and Engineering 
Research Council of Canada for financial support.

\appendix

\section{Covariant canonical formalism on the cotangent bundle}
\label{sc:apx}

It is textbook knowledge \cite{Arnold:book} that the cotangent bundle ${\coTB}$
has the natural
structure of a phase space, i.e., it is possesses a symplectic form ${\tens{\Omega}}$ 
which defines the Poisson bracket ${\{\;,\;\}}$. For a base manifold ${M}$ which is
equipped with an additional geometric structure, it can be useful to express phase-space
quantities and operations with the help of quantities and operations on the base manifold.
In this Appendix we shortly review such a procedure.\footnote{%
Similarly to the main text we type tensors in bold.  
Optionally, we write here the tensors with \emph{abstract indices} 
\cite{PenroseRindler:book,Wald:book1984}
which help to indicate tensorial operations as, for example, contraction.
However, the abstract indices do not refer to any particular choice of coordinates.
We use small latin letters for base manifold indices (for tensors from ${\mathbf{T} M}$),
but we do not introduce indices for the phase space tensors (tensors from ${\mathbf{T}\phsp}$).
We assume implicitly the tensor product, i.e., ${\tens{a}\tens{b}=\tens{a}\otimes\tens{b}}$.}

We call functions on the phase space ${\phsp=\coTB}$ observables, and
we write ${F(x,\tens{u})}$ to emphasize 
the dependence of ${F}$ on the configuration variable ${x\in M}$ and 
on the momentum ${\tens{u}\in\mathbf{T}_x^{*}M}$.

For the base manifold $M$ with a (torsion-free) covariant derivative ${\covd}$ 
(in our case the spacetime manifold with the metric connection), 
it is possible to introduce the covariant derivative of an observable 
${F(x,\tens{u})}$ \emph{in the horizontal (configurational) direction} of the phase space ${\phsp}$.
For any base manifold vector ${\tens{l}\in\mathbf{T}M}$ we define
\begin{equation}\label{pscovddef}
  \tens{l}^{e}\covd_{\!e}F(x,\tens{u})=\frac{d}{d\alpha}F(x(\alpha),\tens{u}(\alpha))\Big|_{\alpha=0}\;,
\end{equation}
where ${x(\alpha)}$ is a curve starting from ${x}$ with tangent vector ${\tens{l}}$,
and ${\tens{u}(\alpha)}$ is parallel transport of ${\tens{u}}$ along ${x(\alpha)}$.

For a 1-form ${\tens{p}\in\mathbf{T}_x^{*}M}$ we can also introduce 
the derivative of an observable ${F(x,\tens{u})}$ \emph{in the vertical (momentum) direction}
\begin{equation}
   \tens{p}_e \pder^e F(x,\tens{u}) = \frac{d}{d\alpha}F(x,\tens{u}+\alpha\tens{p})\Big|_{\alpha=0}\;.
\end{equation}
Thanks to the linearity of ${\mathbf{T}_x^{*}M}$, this derivative is independent of 
any additional geometrical structure.

Derivatives ${\tens{l}^e\covd_{\!e}F}$ and ${\tens{p}_e\pder^{e}F}$
are derivative operators on ${\phsp}$ and as such 
they define vector fields on ${\phsp}$,
which we denote\footnote{%
We could be more explicit and write them as 
${\tens{l}^e\frac{\covd_{\!e}}{\tens{\partial} x}}$
and ${\tens{p}_e\frac{\pder^{e}}{\tens{\partial}\tens{u}}}$.
Similarly we could write ${\frac{\covd_{\!e}F}{\tens{\partial} x}}$
and ${\frac{\pder^{e}F}{\tens{\partial}\tens{u}}}$ for quantities 
${\covd_{\!e}F}$ and ${\pder^e F}$ introduced below.
However, we use such an explicit notation only for the mixed tensor fields
${\frac{\covd_{\!e}}{\tens{\partial} x}}$
and ${\frac{\pder^{e}}{\tens{\partial}\tens{u}}}$ (see below) where
the notation ${\covd_{\!e}}$ and ${\pder^{e}}$ would be too brief.}
${\tens{l}^e\covd_{\!e}}$ and ${\tens{p}_e\pder^{e}}$.
These derivatives and vector fields depend ultralocally on the base 
manifold quantities ${\tens{l}}$ and ${\tens{p}}$
respectively, and we  can thus introduce differentials
${\covd_{\!e}F}$ and ${\pder^e F}$ and mixed tensor quantities 
${\frac{\covd_{\!e}}{\tens{\partial}x}}$ and 
${\frac{\pder^e}{\tens{\partial}\tens{u}}}$
by `tearing off' ${\tens{l}}$ and ${\tens{p}}$,
respectively, and by `tearing off' the function ${F}$.

Clearly, ${\covd_{\!e}F}$ is the covariant derivative of the observable ${F(x,\tens{u})}$
which `ignores' the momentum ${\tens{u}}$ leaving it covariantly constant.
On the other side, the derivative ${\pder^e F}$ `ignores' the configuration 
variable ${x}$. 

For an observable ${F(x,\tens{u})}$ given by a contraction 
of a spacetime tensor field ${\tens{f}(x)}$ with several momenta ${\tens{u}}$,
\begin{equation}\label{fisfuu}
  F(x,\tens{u}) = \tens{f}^{abc\dots}(x)\,\tens{u}_a\tens{u}_b\tens{u}_c\dots\;,
\end{equation}
the covariant derivative reduces to the standard base manifold covariant derivative 
\begin{equation}\label{covdfuu}
  \covd_{\!e}F(x,\tens{u}) = \covd_{\!e}\tens{f}^{abc\dots}(x)\;
  \tens{u}_a\tens{u}_b\tens{u}_c\dots\;.
\end{equation}
The momentum derivative leaves ${\tens{f}}$ intact
\begin{equation}\label{pderfuu}
\begin{split}
  \pder^{e}F(x,\tens{u}) 
    &=\tens{f}^{ebc\dots}(x)\,\tens{u}_b\tens{u}_c\cdots
    +\tens{f}^{aec\dots}(x)\,\tens{u}_a\tens{u}_c\cdots\\
    &\qquad+\tens{f}^{abe\dots}(x)\,\tens{u}_a\tens{u}_b\cdots+\dots\;.
\end{split}\raisetag{3ex}    
\end{equation}
A general phase space observable can then be written
as a (infinite) sum of terms of this type.

The mixed tensor ${\frac{\covd_{\!e}}{\tens{\partial}x}}$ 
is a vector field on the phase space (from ${\mathbf{T}\phsp}$) and a 1-form
on the base manifold (from ${\coTB}$). It is actually the horizontal lift from
${\mathbf{T}M}$ to ${\mathbf{T}\phsp}$ corresponding to the covariant derivative ${\covd}$. 
The mixed tensor 
${\frac{\pder^e}{\tens{\partial}\tens{u}}}$ is a vector field on the phase space
(from ${\mathbf{T}\phsp}$) and a vector field on the base manifold (from ${\mathbf{T}M}$).
It gives a natural identification of the cotangent fiber ${\mathbf{T}_x^{*}M}$ with
its vertical tangent space ${\mathbf{T}\mathbf{T}_x^{*}M}$.

The inverse symplectic form ${\tens{\Omega}^{-1}}$ and the Poisson bracket can be written as
\begin{equation}\label{invSymplStr}
  \tens{\Omega}^{-1} = 
   \frac{\covd_{\!e}}{\tens{\partial}x} \frac{\pder^e}{\tens{\partial}\tens{u}}
   -\frac{\pder^e}{\tens{\partial}\tens{u}} \frac{\covd_{\!e}}{\tens{\partial}x}
\end{equation}
and 
\begin{equation}\label{PB}
  \{A,B\} = \covd_{\!e}A\, \pder^e B - \pder^e A\, \covd_{\!e} B\;.
\end{equation}
They do not depend on a choice of the covariant derivative. Indeed, if we choose
another torsion-free covariant derivative ${\tilde{\covd}}$ on ${M}$, which can be
done by specifying the `difference' tensor ${\tens{\Gamma}^{\,b}_{ac}}$,
\begin{equation}
  \tilde{\covd}_{\!a}\tens{a}^b = \covd_{\!a}\tens{a}^b + \tens{\Gamma}^{\,b}_{ac}\,\tens{a}^c\;,
\end{equation}
the induced covariant derivative of the phase space observables
transforms as
\begin{equation}
  \tilde{\covd}_{\!a}F(x,\tens{u})=\covd_{\!a}F(x,\tens{u}) 
  +\tens{u}_{e}\,\tens{\Gamma}^{\,e}_{ac}(x)\,\pder^{c}F(x,\tens{u})\;.
\end{equation}
Substituting this into \eqref{PB} and using the symmetry
${\tens{\Gamma}^{\,b}_{ac}=\tens{\Gamma}^{\,b}_{ca}}$ we find that 
\begin{equation}
  \{A,B\} = \tilde\covd_{\!e}A\, \pder^e B - \pder^e A\, \tilde\covd_{\!e} B\;,
\end{equation}
i.e., the Poisson bracket is independent of the choice of the covariant derivative.
The argument for the symplectic structure is similar.

The Poisson bracket of an observable of type \eqref{fisfuu} with an observable ${p}$
linear in momenta ${\tens{u}}$,
\begin{equation}
  p(x,\tens{u})=\tens{l}^c(x)\,\tens{u}_c\;,
\end{equation}
leads, with help of \eqref{covdfuu} and \eqref{pderfuu}, to the Lie derivative:
\begin{equation}\label{PBlinobs}
\begin{split}
  &\{F,p\}  
  = \tens{l}^e\covd_{\!e}F - \pder^e F\, (\covd_{\!e}\tens{l}^c)\,\tens{u}_c\\
  &\;=\bigl(\tens{l}^e\covd_{\!e}\tens{f}^{ab\dots}
  - \tens{f}^{eb\dots} \covd_{\!e}\tens{l}^a
  - \tens{f}^{ae\dots} \covd_{\!e}\tens{l}^b
  - \dots\bigr)
  \,\tens{u}_a\tens{u}_b\cdots\\
  &\;=\bigl(\lied_{\tens{l}}\tens{f}^{ab\dots}\bigr)\,\tens{u}_a\tens{u}_b\cdots\;
  \equiv\lied_{\tens{l}}F\;.
\end{split}\raisetag{3ex}
\end{equation}
Here ${\lied_{\tens{l}}\tens{f}}$ is the standard Lie derivative on ${M}$
along the vector field ${\tens{l}}$. 
The last equality then defines the generalized 
Lie derivative ${\lied_{\tens{l}}F}$
of the phase space observable ${F}$
along the base manifold vector field ${\tens{l}}$
which effectively `ignores' the
dependence of ${F}$ on the momentum ${\tens{u}}$.
It can be extended to general phase space observables 
by linearity. It can be also defined similarly  to \eqref{pscovddef}
with ${\tens{u}(\alpha)=\phi_\alpha\tens{u}}$ given by a flow ${\phi_\alpha}$ induced
by the vector field ${\tens{l}}$ acting on ${\tens{u}}$.
${\lied_{\tens{l}}F}$ can be also viewed as the derivative
of the observable ${F}$ along the vector field ${\lied_{\tens{l}}}$
on ${\coTB}$ which is called the complete lift of the vector
field ${\tens{l}}$ on~${M}$.

Clearly, the Poisson bracket with the Hamiltonian \eqref{Ham}
leads to the covariant derivative along the ${\tens{u}}$ direction:
\begin{equation}\label{PBHam}
\begin{split}
  &\{F,H\} = \tens{u}^e\covd_{\!e}F\;.
\end{split}
\end{equation}

Despite the fact that we do not need them in the main text, let us
introduce for completeness the mixed tensor fields ${\tens{D}^{e}x}$
and ${\covd\tens{u}_{e}}$ dual to
${\frac{\covd_{\!e}}{\tens{\partial}x}}$ and 
${\frac{\pder^e}{\tens{\partial}\tens{u}}}$ defined by
\begin{equation}
\begin{gathered}
 \frac{\covd_{\!b}}{\tens{\partial}x} \cdot \tens{D}^{a}x = \tens{\delta}^{a}_{b}\;,\quad
 \frac{\pder^a}{\tens{\partial}\tens{u}} \cdot \covd\tens{u}_{b}= \tens{\delta}^{a}_{b}\;,\\
 \frac{\covd_{\!a}}{\tens{\partial}x} \cdot \covd\tens{u}_{b}= 0\;,\quad
 \frac{\pder^a}{\tens{\partial}\tens{u}} \cdot \tens{D}^{b}x =0\;.
\end{gathered}
\end{equation}\\[0ex]
Here the dot `$\cdot$' indicates the contraction of the phase space tensor indices.

${\tens{D}^{e}x}$ is a vector field on the base manifold ${M}$ and
a \mbox{1-form} on the phase manifold ${\phsp}$. It is actually the
differential of the  bundle projection ${x:\coTB\to M}$.
${\covd\tens{u}_{e}}$ is a 1-form both on the base manifold ${M}$ and
phase space ${\phsp}$.

These phase space `forms' satisfy the completeness relation
\begin{equation}
 \frac{\covd_{\!e}}{\tens{\partial}x}\, \tens{D}^{e}x 
 +\frac{\pder^e}{\tens{\partial}\tens{u}}\, \covd\tens{u}_{e}= \tens{\delta}\;,
\end{equation}
with ${\tens{\delta}}$ being the identity  tensor on ${\mathbf{T}\phsp}$.
The symplectic structure ${\tens{\Omega}}$ can be written as
\begin{equation}\label{SymplStr}
  \tens{\Omega} = \tens{D}^{e}x\,\covd\tens{u}_{e}-\covd\tens{u}_{e}\,\tens{D}^{e}x\;.
\end{equation}

Finally, if we choose the coordinate derivative ${\coord}$ 
associated with a coordinate system ${x^a}$ on ${M}$,
\begin{equation}
  \coord \cv_{x^a}=0\;,\quad\coord\grad x^a=0\;,
\end{equation}
instead of the covariant derivative ${\covd}$, the relations
\eqref{SymplStr}, \eqref{invSymplStr}, and \eqref{PB} reduce
to the standard relations in terms of the canonical coordinates
${x^a,u_b}$ on ${\phsp}$, namely
\begin{equation}\label{SymplStrincoor}
\begin{gathered}
  \tens{\Omega} =\grad{x^e}\, \grad{u_e}-\grad{u_e}\, \grad{x^e}
  \;,\\[1ex]
  \tens{\Omega}^{-1}= \cv_{x^e}\, \cv_{u_e} - \cv_{u_e}\,\cv_{x^e}
  \;,
\end{gathered}
\end{equation}
and
\begin{equation}
  \{A,B\}
  =\frac{\partial A}{\partial x^e}\,\frac{\partial B}{\partial u_e}
  -\frac{\partial A}{\partial u_e}\,\frac{\partial B}{\partial x^e}\;.
\end{equation}
All coordinate vectors and 1-forms in \eqref{SymplStrincoor} 
live on the phase space ${\phsp}$.



\begin{thebibliography}{}


\bibitem{PKVK} D.N.~Page, D.~Kubiz\v n\'ak, M.~Vasudevan, and
P.~Krtou\v{s}, Phys.\ Rev.\ Lett.\ {\bf 98}, 061102 (2007),
arXiv:hep-th/0611083.

\bibitem{KKPF} P.~Krtou\v{s}, D.~Kubiz\v n\'ak, D.N.~Page, and
V.P.~Frolov, J. High Energy Phys. 02 (2007) 004, arXiv:hep-th/0612029.

\bibitem {Schw} K.~Schwarzschild,  Sitzungsber. deutsch. Akad. Wiss.
Berlin, Kl. Math. Phys. Technik, 189 (1916).

\bibitem{Tang} F.R.~Tangherlini, Nuovo Cimento\ {\bf 27}, 636 (1963).

\bibitem{Kerr} R.P.~Kerr, Phys.\ Rev.\ Lett.\ {\bf 11}, 237 (1963).

\bibitem{MP} R.C.~Myers and M.J.~Perry, Ann. Phys. (N.Y.) {\bf 172},
304 (1986).

\bibitem{HHT} S.W.~Hawking, C.J.~Hunter, and M.M.~Taylor-Robinson,
Phys.\ Rev.\ D\  {\bf 59}, 064005 (1999), arXiv:hep-th/9811056.

\bibitem{Carter1} B.~Carter, Phys.\ Lett.\ {\bf 26A}, 399 (1968).

\bibitem{Carter2} B.~Carter, Commun.\ Math.\ Phys.\ {\bf 10}, 280
(1968).

\bibitem{GLPP1} G.W.~Gibbons, H.~L\"u, D.N.~Page and C.N.~Pope, J.\
Geom.\ Phys.\ {\bf 53}, 49 (2005), arXiv:hep-th/0404008.

\bibitem{GLPP2} G.W.~Gibbons, H.~L\"u, D.N.~Page and C.N.~Pope, Phys.\
Rev.\ Lett.\  {\bf 93}, 171102 (2004), arXiv:hep-th/0409155.

\bibitem{CLP} W.~Chen, H.~L\"u, and C.N.~Pope, Class.\ Quant.\ Grav.\
{\bf 23}, 5323 (2006), arXiv:hep-th/0614125.

\bibitem{NUT} E.~Newman, L.~Tamburino, and T.~Unti, J.\ Math.\ Phys.\
(N.Y.)\ {\bf 4}, 915 (1963).

\bibitem{KNAcurv} N.~Hamamoto, T.~Houri, T.~Oota, and Y.~Yasui, J.
Phys. {\bf A40}, F177 (2007), arXiv:hep-th/0611285.

\bibitem{KF} D.~Kubiz\v n\'ak and V.P.~Frolov, Class.\ Quant.\ Grav.\
{\bf 24}, F1 (2007), arXiv:gr-qc/0610144.

\bibitem{Arnold:book} V.I.~Arnol'd, {\it Mathematical Methods of
Classical Mechanics} (Springer-Verlag, Berlin, 1978).

\bibitem{Kozlov} V.V.~Kozlov, Usp.\ Mat.\ Nauk {\bf 38:1}, 3 (1983)
[Russ.\ Math.\ Surv.\ {\bf 38:1}, 1 (1983)].

%

\bibitem{FKK} V.P.~Frolov, P.~Krtou\v{s}, and D.~Kubiz\v n\'ak, J.\
High Energy Phys.\ {\bf 02}, 005 (2007), arXiv:hep-th/0611245.

\bibitem{Benenti} S.~Benenti and M.~Francaviglia, Gen.\ Rel.\ Grav.\
{\bf 10}, 79 (1979).

\bibitem{PenroseRindler:book} R. Penrose and W. Rindler, {\em Spinors
and Space-Time} (Cambridge University Press, Cambridge, England, 1984,
1986).

\bibitem{Wald:book1984} R.~M. Wald, {\em General Relativity} (The
University of Chicago Press, Chicago and London, 1984).

\end{thebibliography}

\end{document}